\documentclass[11pt]{article}
\usepackage{amssymb,epsfig}
\usepackage{amsthm}
\usepackage[numbers]{natbib}
\def\citet{\cite}
\usepackage{etoolbox}
   \csname @addtoreset\endcsname{equation}{section}
    \csname @addtoreset\endcsname{figure}{section}
    \csname @addtoreset\endcsname{table}{section}

\newcounter{thanksnum}
\def\thanksnumber#1
{\setcounter{thanksnum}{\value{footnote}}\setcounter{footnote}{#1}%
                     \addtocounter{footnote}{-1}\footnotemark
                     \setcounter{footnote}{\value{thanksnum}}}
\def\newtheoremz#1{\@ifnextchar[{\@othmz{#1}}{\@nthmz{#1}}}

\def\@nthmz#1#2{%
\@ifnextchar[{\@xnthmz{#1}{#2}}{\@ynthmz{#1}{#2}}}

\def\@xnthmz#1#2[#3]{\expandafter\@ifdefinable\csname #1\endcsname
{\@definecounter{#1}\@addtoreset{#1}{#3}%
\expandafter\xdef\csname the#1\endcsname{\expandafter\noexpand
  \csname the#3\endcsname \@thmcountersepz \@thmcounterz{#1}}%
\global\@namedef{#1}{\@thmz{#1}{#2}}\global\@namedef{end#1}{\@endtheoremz}}}

\def\@ynthmz#1#2{\expandafter\@ifdefinable\csname #1\endcsname
{\@definecounter{#1}%
\expandafter\xdef\csname the#1\endcsname{\@thmcounterz{#1}}%
\global\@namedef{#1}{\@thm{#1}{#2}}\global\@namedef{end#1}{\@endtheoremz}}}

\def\@othmz#1[#2]#3{\expandafter\@ifdefinable\csname #1\endcsname
  {\global\@namedef{the#1}{\@nameuse{the#2}}%
\global\@namedef{#1}{\@thmz{#2}{#3}}%
\global\@namedef{end#1}{\@endtheoremz}}}

\def\@thmz#1#2{\refstepcounter
    {#1}\@ifnextchar[{\@ythmz{#1}{#2}}{\@xthmz{#1}{#2}}}

\def\@xthmz#1#2{\@begintheoremz{#2}{\csname the#1\endcsname}\ignorespaces}
\def\@ythmz#1#2[#3]{\@opargbegintheoremz{#2}{\csname
       the#1\endcsname}{#3}\ignorespaces}

\def\@thmcounterz#1{\noexpand\arabic{#1}}
\def\@thmcountersepz{.}
\def\@begintheoremz#1#2{ \trivlist \item[\hskip \labelsep{\bf #1\ #2}]}
\def\@opargbegintheoremz#1#2#3{ \trivlist
      \item[\hskip \labelsep{\bf #1\ #2\ (#3)}]}
\def\@endtheoremz{\endtrivlist}

\newtheorem{theorem}{Theorem}
\newtheorem{lemma}{Lemma}
\newtheorem{corollary}{Corollary}
\newtheorem{definition}{Definition}
\newtheorem{remark}{Remark}

\def\e{\varepsilon}

\def\defi{\stackrel{{\scriptscriptstyle \Delta}}{=}}

\def\d{\delta}
\def\o{\omega}
\def\O{\Omega}

\def\Y{{\cal Y}}

\def\w{\widehat}
\def\Ind{{\,\rm Ind\,}}
\def\Ind{{\mathbb{I}}}

\def\mes{{\rm mes\,}}

\def\essinf{\mathop{\rm ess\, inf}}
\def\const{{\rm const\,}}

\def\Var{{\rm Var\,}}

\def\R{{\bf R}}
\def\E{{\bf E}}
\def\P{{\bf P}}

\def\Z{{\cal Z}}

\def\J{{\cal J}}

\def\s{\delta}
\def\g{\gamma}
\def\C{{\bf C}}

\def\ww{\widetilde}
\def\X{{\cal X}}
\def\t{\theta}
\def\oo{\bar}
\def\s{\sigma}

\newcommand{\be}{\begin{equation}}
\newcommand{\ee}{\end{equation}}
\newcommand{\bd}{\begin{displaymath}}
\newcommand{\ed}{\end{displaymath}}
\newcommand{\ba}{\begin{array}{ll}}
\newcommand{\ea}{\end{array}}
\newcommand{\baa}{\begin{eqnarray}}
\newcommand{\eaa}{\end{eqnarray}}
\newcommand{\baaa}{\begin{eqnarray*}}
\newcommand{\eaaa}{\end{eqnarray*}}
\font\sm=cmr10

\def\oo{\bar}

\def\LBL{{\scriptscriptstyle LBL}}
\def\BL{{\scriptscriptstyle BL}}

\def\Q{{\cal Q}}

\def\sinc{{\rm sinc\,}}
\def\ew{\left(e^{i\o}\right)}
\def\T{{\mathbb{T}}}
\def\ZZ{{\mathbb{Z}}}

\def\TT{{\cal T}}
\def\XN{\ell_2^{-,\LBL}(I)}
\def\XNil{\ell_2^{-,\LBL}}

\def\BN{{\mathbb{B}}(I)}
\def\BNil{{\mathbb{B}}}

\def\BNil{{\mathbb{B}}}
\def\BN{{{\bf X}_\BL}(I)}
\def\BNil{{\X}}

\def\YYN{{\ell_2}}
\def\om{\o_0}
\date{Submitted: January 22, 2016. Revised: June 22, 2016 }
\title{On  detecting and quantification of randomness  for one-sided sequences
}\author{
Nikolai Dokuchaev\index{Corresponding author;}\\
 {\sm Department of Mathematics \& Statistics, Curtin
University,}\\ {\sm  GPO Box U1987, Perth, 6845 Western Australia}\index{\sm tel. 61892663144; fax
618 92663197; email N.Dokuchaev@curtin.edu.au} }
\begin{document}
\def\break{}%
\def\brea{}
\def\breakk{}
\maketitle
\begin{abstract}
The paper studies  discrete time processes and their predictability and randomness
in deterministic pathwise setting, without using  probabilistic assumptions on the ensemble.
We suggest
some approaches to quantification of randomness based on frequency analysis of two-sided and one-sided sequences. In addition, the paper suggests an extension of the notion
of bandlimitiness on one-sided sequences and a  procedure allowing to represent an one-sided sequence as a sum of left-bandlimited and predictable sequences and a non-reducible noise.
\end{abstract}
\section{Introduction}
The paper studies  discrete time processes and their predictability and randomness
in deterministic pathwise setting, without using  probabilistic assumptions on the ensemble.
\par
Understanding of the pathwise randomness leads to many applications in Monte-Carlo methods, cryptography, and control systems.
There are many classical works devoted to the concept of pathwise randomness and the problem of distinguishability
 of random sequences; see the references in \cite{LV,Dow}. \index{Li and Vitanyi
(1993) and  Downey (2004).} In particular, the approach from \
 Borel (1909) \cite{Bor}  , Mises (1919) \cite{Mis}  , Church (1940) \cite{Ch}   was   based  on
 limits of the sampling proportions of zeros in the binary sequences and subsequences; Kolmogorov
(1965) \cite{Ko65} and Loveland (1966) \cite{Lo} developed a different  concept of the algorithmic randomness and compressibility;
Schnorr (1971) \cite{Sch} suggested approach based on predicability and  martingale properties. So far, the exiting theory is devoted to the problem of distinguishability
 of random sequences  and does not consider the problem  of quantification of the degree of randomness.
This paper studies randomness in the sense of
the pathwise  predicability and attempts to develop an approach for  quantification and separation of the ``noise" for the
sequences that are deemed to be random. The estimation of the degree on randomness is  a difficult  problem, since the task of
detecting the randomness  is  nontrivial itself.
\par
The paper  investigates randomness  and noise for the
sequences in a more special setting originated from the linear filtering and prediction of stochastic processes
rather than algorithmic randomness in the sprit of   Downey (2004).  We suggest exploring  the following straightforward pathwise
criterion:  a class of sequences that is
predictable or such that its missing value can be recovered without error from observations of remaining values
 is assumed to consist of  non-random
sequences.

For stationary discrete time processes,  there
is a criterion of predictability and recoverability in the frequency domain setting
given by the classical minimality criterion  \cite{Ko41},
Theorem 24, and the Szeg\"o-Kolmogorov theorem; see \cite{Sz,V}\index{
Szeg\"o (1920, 1921), Verblunsky  (1936), Kolmogorov (1941),\index{
Theorem 2}} and recent literature reviews in \cite{Bin,S}.\index{ Simon (2011) and Bingham
(2012).} By this theorem,  a stationary process  is
predictable if its spectral density is vanishing with a certain rate at a point of the
unit circle $\{z\in\C:\ |z|=1\}$. In particular, it holds if the spectral
density vanishes on an arc of
the unit circle, i.e., the  process is bandlimited.  There are many works devoted to smoothing in frequency domain and sampling; see, e.g.,
\cite{Alem,rema, F94, FKR, Leef, TH,W} and the bibliography here.
  \index{\citet{jerry}, \citet{pollock}, \citet{PFG}, \citet{PFG1}, \citet{AU}.}  In \cite{D10,D12a,D12b,D12c,D12d,D16}, predictability  was
readdressed in the deterministic setting for two-sided sequences for
with Z-transform vanishing in a point on $\T$, and some predictors
were suggested.  These results were based on frequency characteristics of the entire
two-sided sequences, since the properties of the Z-transforms were used.   Application of the two-sided Z-transform requires to select some
past time at the middle of the time interval of the observations as the zero point for a model of  the two-sided sequence; this could be inconvenient.
 In many applications, it is more convenient
 to represent data flow
as one-sided sequences such that $x(t)$  represents outdated observations with diminishing significance as
$t\to -\infty$.
This leads to the analysis of the  one-sided sequences directed backward to the past. However,
the straightforward application of the one-sided
Z-transform to the one-sided sequences  does not generate Z-transform vanishing on a part of the unit circle even for a band-limited underlying  sequence.

The paper suggests
some approaches to quantification of randomness based on frequency analysis of two-sided and one-sided sequences.
In addition, the paper suggests an extension of the notion
of bandlimitiness on one-sided sequences and a  procedure allowing to represent an one-sided sequence as a sum of left-bandlimited and predictable sequences and a non-reducible noise.

\section{Definitions and background}
We use notation $\sinc(x)=\sin(x)/x$ and  $\T=\{z\in\C:\  |z|=1\}$, and we denote by $\ZZ$  the set of all integers.
\par
 For a Hilbert space $H$, we denote by $(\cdot,\cdot)_{H}$ the
corresponding inner product. We denote by $L_2(D)$ the usual Hilbert space of complex valued
square integrable functions $x:D\to\C$, where $D$ is an interval  in $\R$.

Let $\tau\in\ZZ\cup\{+\infty\}$ and $\t<\tau$; the case where
$\t=-\infty$  is not excluded. We denote by $\ell_r(\t,\tau)$ the
Banach space of complex valued sequences $\{x(t)\}_{t=\t}^\tau$
such that
$\|x\|_{\ell_r(\t,\tau)}=\left(\sum_{t=\t}^\tau|x(t)|^r\right)^{1/r}<+\infty$
for $r\in[1,\infty)$ or  $\|x\|_{\ell_\infty(\t,\tau)}=\sup_{t: \t-1<t<\tau+1}|x(t)|<+\infty$
for $r=+\infty$.
\par
Let $\ell_r=\ell_r(-\infty,+\infty)$ and
$\ell_r^-=\ell_r(-\infty,0)$.
\par
For  $x\in \ell_1$ or $x\in \ell_2$, we denote by $X=\Z x$ the
Z-transform  \baaa X(z)=\sum_{t=-\infty}^{\infty}x(t)z^{-t},\quad
z\in\T. \eaaa Respectively, the inverse Z-transform  $x=\Z^{-1}X$ is
defined as \baaa x(t)=\frac{1}{2\pi}\int_{-\pi}^\pi
X\left(e^{i\o}\right) e^{i\o t}d\o, \quad t=0,\pm 1,\pm 2,....\eaaa
If $x\in \ell_2$, then $X|_\T$ is defined as an element of
$L_2(\T)$.

\par
\index{Let $\Y_N$ be the Hilbert space of sequences $\{y_k\}_{k=
-N}^{N}\subset\C$ provided with the $\ell_2$-norm, i.e.,
$\|y\|_{\Y_N}=\left(\sum_{k\in \ZZ_N }|y_k|^2\right)^{1/2}<+\infty$.}

\par

For a set $I\subset (-\pi,\pi]$, we denote $I^c=(-\pi,\pi]\backslash I$.

Let $\J$ be the set of all $I\subset (-\pi,\pi]$ such  that the set $\{e^{i\o}\}_{\o\in I}$ is a connected arc and $I^c\neq \emptyset$.

For any $I\in\J$,  we  denote by $\o_I$ the middle point $e^{i\o_I}$ of the arc $\{e^{i\o}\}_{\o\in I}$.

We denote by $\BN$  the set of all mappings $X:\T\to\C$ such
that $X\ew \in L_2(-\pi,\pi)$ and $X\ew =0$ for $\o\notin I$. We will call  the corresponding processes $x=\Z^{-1}X$
{\em band-limited}. Let $\ell_2^\BL(I)=\{x\in\ell_2:\ X=\Z x\in \BN\}$.
\begin{definition}  Assume that there exists $I\in\J$ such that
   $x\in\ell_2^-$ represents the  trace of a band-limited process $x_{\BL}\in\ell_2^\BL(I)$ with the spectrum on $I$, i.e., $x(t)=x_{\BL}(t)$ for $t\le 0$, and $X_{\BL}=\Z x_{\BL}\in \BN$. We call the process  $x$
\underline{ left band-limited} with the spectrum on $I$.
\end{definition}

 Let $\XN$ be the subset of $\ell_2^-$ consisting of semi-infinite sequences
$\{x(t)\}_{t\le 0}$ such that $x(t)=(\Z^{-1}
X)(t)$ for $t\le 0$ for some $X \in \BN $.

\section{Quantification of randomness for two-sided sequences}  \label{Sec2Sided}
Let us discuss first a straightforward approach where noise is associated with the high-frequency component.
Consider a
sequence $x\in\ell_{2}$ that does not feature predicability described in Lemma \ref{lemmaPred}.
Let
\baaa
X=\Z X,\qquad Y_\BL|_{\T}=(\Ind_I X)|_{\T},\qquad y_\BL=\Z^{-1} Y_\BL
\eaaa for some given $I\in\J$. Here $\Ind$ is the indicator function, i.e.,
$\Ind_I\ew=1$ if $\o\in I$ and $\Ind_I\ew=0$ if $\o\in I^c=(-\pi,\pi]\backslash I$.
In many applications, it is acceptable to deem the process $n_\BL(t)=x(t)-y_\BL(t)$  with $I=I_0=(-\O,\O)$, where $\O\in(0,\pi)$, to be a noise
accompanying the systematic movement $y_\BL(t)$.
However, estimation
of $n_\BL(t)$ will not help to quantify the randomness of $x$, since
\baa
\|n_\BL\|_{\ell_2}=\|x-y_\BL\|_{\ell_2}\to 0\quad \hbox{as}\quad \mes(-\pi,\pi]\setminus I)\to 0\quad \hbox{for any}\quad x\in\ell_2.
\label{2s}\eaa
In  addition, $n_\BL$  is also a predictable  band-limited process,
\baaa
n_\BL=x-y_\BL=\Z^{-1}(\Ind_{I^c} \Z x)\in \ell_2^\BL(I^c),
\label{2}
\eaaa
 In particular, this means that
any two-sided sequence  $x\in \ell_2$ can be represented as
 as a sum \baa
x=y_\BL+ n_\BL, \quad y_\BL\in \ell_2^\BL (I),\quad n_\BL\in\ell_2^\BL (I^c),
\label{xxy}\eaa
  i.e., as   a sum of two two-sided  band-limited predictable  in the sense of Lemma \ref{lemmaPred} sequences, and that can be done with any choice of $I$.
\index{In addition,   $(\w x_\BL,y_\BL)_{\ell_2}=0$ since $( x_\BL,y_\BL)_{\ell_2}=0$ for any
$x_\BL\in \ell_2^\BL(I)$ and $y_\BL\in \ell_2^\BL(I^c)$.}
This also does not lead to possibility to detect   and quantify randomness.

We suggest a different approach.
We will show below a meaningful  quantification of the randomness of $x\in \ell_2$ can be achieved with the value
\baa
\s(x)=\essinf_{\o\in(-\pi,\pi]}|X\ew|,\quad X= \Z x.
\label{sig}
\eaa

\subsection{Randomness as a measure of non-predictability}
\subsubsection*{Some predictability results for two-sided sequences}
Two-sided band-limited sequences are predictable in the following sense.
\begin{theorem}\label{lemmaPred} \begin{enumerate}
\item
Let $\X\subset \ell_2^\BL(I)$ be a bounded set. Then, for any $\e>0$, there exists a mapping $\w k(\cdot):\ZZ\to\R$
 such that  $\sup_{t\in \ZZ}\|x(t)-\w x(t)\|\le \e$ for all $x\in\X$
for $\w x(t)\defi
e^{i\o_I t} \sum_{s\le t-1} \w k(t-s)e^{-i\o_I s} x(s)$.
\item Let $\J_1\subset\J$ be a set of $I$ such that $\sup_{I\in J_1}\mes(I)<2\pi$.  Let $\X\subset \cup_{I\in \J_1} \ell_2^\BL(I)$ be a bounded set in $\ell_2$. Then, for any $\e>0$, there exists a mapping $\w k(\cdot):\ZZ\to\R$
 such that  $\sup_{t\in \ZZ}\|x(t)-\w x(t)\|\le \e$ for all $x\in\X$
for $\w x(t)\defi
e^{i\o_I t} \sum_{s\le t-1} \w k(t-s)e^{-i\o_I s} x(s)$.
\item
Let $\J_1$ be the set of $I\in\J$ such that $\sup_{I\in J}\mes(I)<2\pi$, and  $\X\subset \cup_{I\in \J_1} \ell_2^\BL(I)$ be a bounded set in $\ell_2$ such
that $\sum_{t\le \tau}|x(t)|^2\to 0$ as $\tau\to +\infty$ uniformly over $x\in \X$. Then, for any $\e>0$, there exists $\tau<0$ and a mapping $\w k(\cdot):\ZZ\to\R$
 such that  $\sup_{t\ge 1}\|x(t)-\w x(t)\|\le \e$ for all $x\in\X$
for $\w x(t)\defi
e^{i\o_I t} \sum_{s=\tau}^{t-1} \w k(t-s)e^{-i\o_I s} x(s)$.
\end{enumerate}
\end{theorem}

Theorem \ref{lemmaPred}(iii) states  that some predicability  based on finite sets of  observations also can be achieved if
we relax  predicability requirement to cover times $t\ge 1$ only;  this would be a weaker version of predicability comparing with the one described in Theorem \ref{lemmaPred} (ii).

Some versions of this Theorem and some examples of predictable classes can be found in \cite{D12a,D12b}.

In addition, it appears that the spectrum supporting sets $I$ can be estimated from the
set of observations $\{x(s)\}_{s\le \tau}$ for any $\tau<0$. More precisely, the following theorem holds.
\begin{theorem}\label{ThO}
Let $\X\subset  \ell_2$ be a  set such that if $x\in\X$ then $x\in \ell_2^\BL(I)$ for some $I=I(x)\in\J$,
and that $\nu\defi 2\pi-\sup_{x\in \X}\mes(I(x))>0$. Let $\w\nu=\nu/3$. Then, for  any $\tau<0$,
 there exists a mapping $F:\ell_2(-\infty,\tau)\to (-\pi,\pi]$ such that,
 for $\w\o_c=F(x(t)|_{t\le\tau})$, $\T_c
 \subset \{e^{i\o},\ \o\in I^c\}$, where
 \baaa
\T_c=\left\{e^{i(\o+\pi)}:\ \o\in (-\pi,\pi],\ \min_{k=0,\pm 1}|\w\o_c-\o+2k\pi|\le\w\nu\right\}.\eaaa
In other words, if $x\in \X$, then $x\in\ell_2^\BL(\w I)$  and $I\subset \w I$, where \baaa
\w I=\left\{\o\in (-\pi,\pi]: e^{i\o}\notin\T_c\right\}.
\eaaa
\end{theorem}
The set $\w I$ in Theorem \ref{ThO} can be regarded as an estimate of $I$ based on observations of $\{x(t)\}_{t\le \tau}$.

Let $\X\subset \ell_2\cap \ell_1$ be a class of processes such that $\s(x)>0$ for $x\in\X$ and that, for $x\in\X$ and $X=Z x$, for any $m>0$, the
functions $X\ew$ and $|X\ew|^{-1}$ are differentiable in $\o\in\R$ and that
$\sup_{x\in\X}\sup_{\o\in[-\pi,\pi]}|dX\ew/d\o|<+\infty$. For the purpose of the
investigation of the predictability for $x$, this smoothness and assumed   without a loss generality: it is
sufficient to replace $x$ by a faster vanishing processes with the
same predictability properties such that
$x(t)/(1+|t|^m)$, $m\ge 2$. $\s=\min_{\o\in[-\pi,\pi]}|X\ew|>0$

  We want to
 represent each $x\in \X$ as
\baaa
x=y_\BL+n,
\eaaa
where $y_\BL$ is a band-limited predictable process such that the class $\Y=\{y_\BL\}_{x\in\X}$,
is predictable in the sense of Lemma \ref{lemmaPred}. In this case, each
  $n= x-y_\BL$ is a non-predictable (random) noise.

  We suggest the following  restrictions on the choice of $y_\BL$:
  \begin{enumerate}
  \item  \baa
\|X\ew\|_{L_d(-\pi,\pi)}=\|Y_\BL\ew\|_{L_d(-\pi,\pi)}+\|N\ew\|_{L_d(-\pi,\pi)},\quad d=1,+\infty, \label{N}\eaa
 where $Y_\BL=\Z y_\BL$ and  $N=\Z n'$.
  \item
  $n$ does not allow a similar representation $n=y'_\BL+n'$, with a
non-random (predictable)   non-zero $y'_\BL$ such that \baaa
\|N\ew\|_{L_d(-\pi,\pi)}=\|Y'\ew\|_{L_d(-\pi,\pi)}+\|N'\ew\|_{L_d(-\pi,\pi)},\quad d=1,+\infty, \label{N'}\eaaa
where $Y'_\BL=\Z y'_\BL$ and  $N'=\Z n'$.
  \end{enumerate}
It appears that  $n$ featuring these properties exists in some case
 and can be derived
explicitly from $X$. Let us show this.

Let $\om\in(-\pi,\pi]$ be such
that $|X\left(e^{i\o_I }\right)| =\s$, and
let  \baa \g\ew=\frac{\s(x)}{|X\ew|},\quad Y\ew=[1-\g\ew]X\ew,\quad
N\ew=\g\ew X\ew. \label{YN}\eaa  Clearly,
\baaa
X=Y+N, \quad Y\left(e^{i\o_I }\right)=0,\quad |N\ew|\equiv\s(x),
\eaaa
and
  (\ref{N}) holds with $d=1$ and $d=\infty$.   By continuity  of  $X\ew$ and
$|X\ew|^{-1}$, the function $Y\ew$ is also continuous $\o$.

If $Y\ew$ vanishes  fast enough when $\o\to\om$ (see \cite{D12a}), then $y=\Z^{-1}Y$ is predictable; in this case, the set
$\{n\}_{x\in\X}$ can be considered as the set of pathwise noises; therefore, this gives  a quantification  $n$ as  a norm of $n$ or $N$, such as
\baa
\|N\ew\|_{L_1(-\pi,\pi)}=\s(x).
\label{noise}
\eaa

However, it would be too restrictive to require  that the set $\X$ is such that (\ref{YN}) leads to $Y\ew$ that vanishes so fast as $\o\to\om$
that $y_\BL$ is predictable. To overcome this, we suggest to replace (\ref{YN}) by
\baa &&\g_\e\ew=1\quad \hbox{if}\quad |e^{i\o}-e^{i\om}|\le \e,\nonumber\\
&&\g_\e\ew=\frac{\s(x)}{|X\ew|}\quad\hbox{if}\quad |e^{i\o}-e^{i\om}|> \e,\quad
\nonumber \\&& Y_\e\ew=[1-\g\ew]X\ew,\qquad
N_\e\ew=\g_\e\ew X\ew,
\label{g}\eaa
where $\e\to 0$.  In this case,
\baa
x=y_\e+n_\e, \quad y_\e=\Z^{-1}Y_\e\in \ell_2^\BL(I_\e),\quad n_\e=\Z^{-1}N_\e,  \label{Yg}\eaa
where $I_\e=\{\o:\ |e^{i\o}-e^{i\om}|\le \e\}$,
\baa  &&|N_\e\ew|= X\left(e^{i\o}\right),\quad \hbox{if}\quad \o\in I_\e,\nonumber\\
&&|N_\e\ew|=|X\left(e^{i\om}\right)|=\const\quad\hbox{if}\quad  \o\notin I_\e,\quad
\eaa
We regard $n_\e$ as approximation of the noise as $\e\to 0+$.

To justify this description of   the noise,  we have to show that the set of band-limited processes  $\{y_\e\}$ in (\ref{g})-(\ref{Yg}) is predictable in some sense. Theorem \ref{lemmaPred}(i)-(ii) does not ensure predicability of this set, since it requires to know  the values $\om$.
This would require to know $\o_{I_\e}$, which is inconsistent   with the notion of predictability. However, Theorem \ref{ThO}
ensures sufficient estimation of $I_\e$  and $\o_{I_\e}$ based on observations of $\{x(t)\}_{t\le \tau}$; we can take  select  $\o_{I_\e}=\w\o_{c}-\pi$ if $\w\o_c\in (0,\pi]$, and
 $\o_{I_\e}=\w\o_{c}+\pi$ if $\w\o_c\in (-\pi,0]$,  in the notations of Theorem \ref{ThO}.
 This leads to the following two step procedure:
the  set $\{x(s)\}_{\tau<s< t}$ is used for prediction of $x(t)$, and the set $\{x(s)\}_{s\le \tau}$ is used for the
estimation of $\w\o_{I_\e}\approx \o_{I_\e}$. This allows to satisfy conditions
of  Theorem \ref{lemmaPred}(iii).

Therefore, the set of band-limited processes  $\{y_\e\}$ in (\ref{g})-(\ref{Yg}) is predictable in the sense of Theorem \ref{lemmaPred}(iii). This  predictability  covers times $t\ge 1$ only; it is a weaker version of predictability comparing with the one described in Lemma \ref{lemmaPred}(i)-(ii).

Since a norm for $N_\e$ is approaching the norm for $N$, the norm of $N$ can be used for quantification
 of the
randomness of the two-sided sequences.

\index{It can be noted that the ``noise" $n=\Z^{-1}N$ does not have to be a
particularly irregular process. For instance, if $X\ew\equiv 1$ then
$Y\ew\equiv 0$ and $N\ew\equiv X\ew\equiv 1$, and the process
$n=\Z^{-1}N$ does not look particularly irregular. However, as was
mentioned above, this process in not allowed in typical classes  of
predictable processes covered by Definition 1.}

The process  $y_\e=\Z^{-1}Y_\e$ can also be interpreted as an output of a
smoothing filter.

This support the choice of the value $\s(x)$ for the quantification of $x$.
\subsection{Randomness as a measure of recoverability}

By recoverability,  we mean a possibility of constructing a  linear  recovering
operator as described in the definition below.

Note that $X\ew$ is continuous in $\o$ for $x\in\ell_1$, $X=\Z x$.

Let   $\om\in (0,\pi]$ be given. For $\s\ge 0$,  let
$\X_\s=\{x\in\ell_1:\ \min_{\o\in(-\pi,\pi]} |X\ew|=|X\left(e^{i\om }\right)|=\s\}$.

For $m\in\ZZ$, assume that the value $x(m)$ is not observable for $x\in\ell_1$ and that all other values of $x$
are observable. We consider recovering problem for $x(m)$ as finding an estimate
$\ww x(m)=F\left(x|_{t\in\ZZ\setminus\{m\}}\right)$, where $F:\ell_1(-\infty,m-1)\times \ell_1(m+1,+\infty)\to\R$ is some mapping.

\begin{theorem} \label{Threc} For any  estimator $\ww x(m)=F\left(x|_{t\in\ZZ\setminus\{m\}}\right)$, where $F:\ell_1(-\infty,m-1)\times \ell_1(m+1,+\infty)\to\R$ is some mapping,
we have that
\baa
\sup_{x\in\X_\s}|\ww x(m)-x(m)|\ge \s.
\label{optrec}\eaa
In addition, there exists an optimal estimator $\w x(m)=\w F\left(x|_{t\in\ZZ\setminus\{m\}}\right)$, where $\w F:\ell_1(-\infty,m-1)\times \ell_1(m+1,+\infty)\to\R$ is some mapping, such that
\baa
\sup_{x\in\X_\s}|\w x(m)-x(m)|=\s.
\label{opt}
\eaa
\end{theorem}

\par \def\s{\sigma}

This supports again  the choice of the value $\s(x)$ for the quantification of the randomness for $x$.

\index{In  some numerical experiments, we calculated  the the normalized size of the noise, i.e. the value \baa
\frac{2\pi\s(x)}{\|X\ew\|_{L_1(-\pi,\pi)}}.\label{noise2}\eaa
for some historical financial time series (returns of stock prices). The value of
(\ref{noise2}) was found  to be in the range of 0.05-0.6 for
different stocks.
}

\index{; this  was consistent with the volatility range for the prices.}
\index{In another experiment,  we matched value (\ref{noise2}) with
the parameter $\s$ used for simulated autoregressions
$x(t+1)=x(t)+\const\cdot  \Delta +\s \Delta^{1/2} \xi(t),$ where
$\xi(t)$ was modelled as a discrete time white noise with $\E
\xi(t)=0$, $\Var \xi(t)^2=1$. These processes are used for financial
modelling; for instance,  the process $x(t)$ with $\Delta=1/250$ can
represent daily returns for a stock with the price volatility $\s$.
The results were quite interesting; more experiments are planned.}

\section{Separating the noise for one-sided sequences}\label{Sec1Sided}
Unfortunately, representation (\ref{xxy}) does not lead toward a solution of  the predictability problem, since it would require to know the entire sequence $\{x(t)\}_{t=-\infty}^{+\infty}$ to calculate $X=\Z x$.

On the other hand, it is natural to  use one-sided sequences interpreted as available past observations  for predictability problems.
For this, we have to use the notion of left  bandlimitness for
one-sided sequences. We will use a modification of representation (\ref{xxy}) that was stated for two-sided sequences.

For this, we have to of representation   to two-sided sequences. We suggest to replace   the "ideal" projections  $\w x_\BL=\Z^{-1}(\Ind_{I} \Z x)\in \ell_2$ for $x\in\ell_2$ and
$y_\BL=x-\w x_\BL=\Z^{-1}(\Ind_{I^c} \Z x)$  by their "optimal" one-sided substitutes.
\subsection*{Uniqueness of the extrapolation for left band-limited processes}
\begin{lemma}\label{propU} For any $I\in\J$ and
  any $x\in\XN $, there exists an
unique  $x_\BL\in\ell_2^\BL(I)$ such that $x(t)=x_{\BL}(t)$ for $t\le 0$.
\end{lemma}
\par
By Lemma \ref{propU}, the future values $x_\BL(t)|_{t>0}$ of a
band-limited process $x_\BL$,
 are uniquely defined by the trace
$x_\BL(t)|_{t\le 0}$.
  This statement represent a reformulation in the deterministic setting
of  the classical Szeg\"o-Kolmogorov Theorem for stationary Gaussian processes
\citet{Ko65,Sz,Sz1,V}.

\subsection*{Existence  of optimal
band-limited approximation} Let $x\in\ell_2^-$ be a semi-infinite one-sided sequence representing available
historical data, and let $I\in\J$.
\begin{theorem}\label{Th1} There exists an unique optimal solution  $\w x$
of the minimization problem \baa &&\hbox{Minimize}\quad  \sum_{t=-\infty}^0|\w
x(t)-x(t)|^2  \quad\breakk\hbox{over}\quad \w x\in \XN .\label{min} \eaa
\end{theorem}
\par
By  Lemma \ref{lemmaPred},  there exists a unique band-limited process  $x_{\BL}\in\ell_2^\BL (I)$
such that $\w x(t)|_{t\le 0}= x_{\BL}(t)|_{t\le 0}$. This offers a natural way  to extrapolate a left
band-limited solution $\w x\in \ell_2^-$ of problem (\ref{min}) on the future
times  $t>0$.

\index{It can
be interpreted  as the optimal forecast (optimal given $\O$ and
$N$).}
\subsubsection*{The optimal solution}
Let $I\in\J$ be given, and let $\mes(I)=2\O$ for some $\O\in (0,\pi)$.

Let $I_0=(-\O,\O)$, i.e.,  $\o_{I_0}=0$.

For $\o\in[-\pi,\pi)$, let the operator $p_{\o}: \ell_2^-\to\ell_2^-$ be defined as $\oo x(t)=e^{i\o  t}x(t)$ for $\oo x=p_\o x$.

Let the operator $\Q: \YYN\to \XNil(I_0) $ be defined as $\w x=\Q y=\Z^{-1}\w X$, where
\baa
\w X\ew =\sum_{k\in
\ZZ}y_ke^{ik\o\pi/\O}\Ind_{\{|\o|\le\O\}},
\label{wX}\eaa  for the corresponding  $y=\{y_k\}\in \YYN$.
Similarly to the classical sinc representation, we obtain that \baa \w x(t)=\frac{1}{2\pi}
\int_{-\O}^{\O}\left(\sum_{k\in \ZZ}y_k e^{ik\o\pi/\O}\right)e^{i\o
t}d\o\brea=\frac{1}{2\pi}
\sum_{k\in \ZZ}y_k\int_{-\O}^{\O}e^{ik\o\pi/\O+i\o t}d\o\nonumber\\
=\frac{1}{2\pi}\sum_{k\in \ZZ}y_k \frac{e^{ik\pi+i\O t}-
e^{-ik\pi-i\O t}}{ik\pi/\O+it}\brea=\frac{\O}{\pi}\sum_{k\in
\ZZ_N }y_k \sinc(k\pi+\O t)=(\Q y)(t).\label{sinc}\eaa
It follows that the $\Q: \YYN\to \XNil(I_0) $ is actually defined as
\baaa \w x(t)=(\Q y)(t)=\frac{\O}{\pi}\sum_{k\in
\ZZ}y_k \sinc(k\pi+\O t).\label{Qs}\eaaa  Consider the operator  $\Q^*:\XNil(I_0) \to \YYN$  being adjoint to the operator
$\Q:\YYN\to\XNil(I_0)$, i.e., such that
\baa
(\Q^*x)_k=\frac{\O}{\pi}\sum_{t\in\TT}\sinc(k\pi+\O t)x(t).
\label{Q*}\eaa

Consider a
linear bounded non-negatively defined  Hermitian operator $R:\YYN\to \YYN$  defined as
\baaa
R=\Q^*\Q.
\eaaa
Consider operator $P_I=p_{\o_I}\Q R^{-1} \Q^* p_{-\o_I}:\ell_2\to\ell_2^{-,\LBL}(I)$.
\begin{theorem}
\label{ThP}
\begin{itemize}\item[(i)] The operator $R:\YYN\to\YYN$ has a bounded inverse
 operator $R^{-1}:\YYN\to\YYN$.
\item[(ii)] Problem (\ref{min}) has a unique solution
 \baa
\w x=P_Ix.\label{wx}
\eaa
\end{itemize}
\end{theorem}
\begin{theorem}\label{Th1n} For any $I\in\J$, there exists $n_I\in \ell_2^-$ such that $P_In_I=0$ and $n_I\neq 0$.
\end{theorem}
The processes $n_I$ can be considered as the noise component with respect to smooth processes with the spectrum on $I$, for a  given $I\in\J$.
\begin{corollary}
\label{corrxx}
A process $x\in\ell_2^-$ is left-bandlimited with the spectrum $I$ if and only if $x=p_{\o_I}\Q R^{-1} \Q^* p_{-\o_I}x$.
\end{corollary}
\begin{remark}
It can be noted that $\w x=p_{\o_I}\Q \Q^+ p_{-\o_I}x$, where $\Q^+=R^{-1} \Q^*:\ell_2^-\to\YYN$ is a Moore--Penrose pseudoinverse of the operator $\Q:\YYN\to\ell_2^-$.
\end{remark}
Let us elaborate equation (\ref{wx}).  The optimal  process $\w x$ can be expressed as  \baaa \w x(t)=e^{i\o_I t}\frac{\O}{\pi}\sum_{k\in \ZZ }\w y_k \sinc(k\pi+\O t). \label{wxx}\eaaa
Here $\w y=\{\w
y_k\}_{k\in\ZZ}$ is defined as    \baa \w y=R^{-1}\Q p_{-\o_I} x.\label{wy}\eaa The operator $R$ can be represented via a matrix
$R=\{R_{km}\}$, where $k,m\in\ZZ$.  In this
setting, $(Ry)_k=\sum_{k=-\infty}^\infty R_{km}y_m$, and  the components of the matrix $R$  are defined as  \baaa R_{km}=
\frac{\O^2}{\pi^2}\sum_{j=-\infty}^0\sinc(m\pi+\O j)\,\sinc(k\pi+\O j)
.\label{R}\eaaa
Respectively, the components of the vector  $\Q^*x=\{(\Q^*x)_k\}_{k\in\ZZ}$   are defined as
\baa (\Q^*x)_{k}= \frac{\O}{\pi} \sum_{j=-\infty}^0\sinc(k\pi+\O j)x(j) .
\label{r}\eaa

\subsection{A multi-step procedure  for one-sided sequences}\label{subsecMS}
Unfortunately, the approach described in Section \ref{Sec2Sided} does not lead toward a solution of  the predictability problem, since it would require to know the entire sequence $\{x(t)\}_{t=-\infty}^{+\infty}$ to calculate
$X=\Z x$ and quantitative characteristics suggested in Section \ref{Sec2Sided}.

On the other hand, it is natural to  use one-sided sequences interpreted as available past observations  for predictability problems.
In this case, we have to use the notion of left  bandlimitness for
one-sided sequences. We will use a modification of representation (\ref{xxy}) that was stated for two-sided sequences.

For this, we suggest to replace   the "ideal" projections  $\w x_\BL=\Z^{-1}(\Ind_{I} \Z x)\in \ell_2$ for $x\in\ell_2$
by their "optimal" one-sided substitutes
$\w x=P_I x\in\ell_2^-$;   this substitution is  optimal on $\{t\le 0\}$ in the sense of optimization problem (\ref{min}).
Unfortunately,  it may happen that
\baaa
x-\w x\notin \ell_2^{-,\LBL}(I^c).
\eaaa

For this, we suggest to replace   the "ideal" projections  $\w x_\BL=\Z^{-1}(\Ind_{I} \Z x)\in \ell_2$ for $x\in\ell_2$ and
$y_\BL=x-\w x_\BL=\Z^{-1}(\Ind_{I^c} \Z x)$  by their "optimal" one-sided substitutes
$\w x=P_I x\in\ell_2^-$ and $\w y=P_{I^c} (x-\w x)\in\ell_2^-$;   this substitution is  optimal on $\{t\le 0\}$ in the sense of optimization problem (\ref{min}).
Unfortunately, \index{an analog of the connection between  (\ref{1}) and (\ref{2}) is not valid for onesided sequences,
i.e.,} it may happen that
\baaa
\w y=P_{I^c} (x-\w x)\notin \ell_2^{-,\LBL}(I^c).
\eaaa

We suggest a  multi-step procedure that to deal with this complication.

Assume that we observe a semi-infinite one-sided sequence $\{x(t)\}_{t\le 0}\in \ell_2^-$.

Consider a sequence of sets $\{I_k\}_{k=0,1,2,..}\subset\J$, with the corresponding middle
points $\o_k\in I_k$.
Further, let us consider the following  sequences of elements of  $\ell_2^-$:
\begin{itemize}
\item Set \baaa
x_0=x,\qquad \w x_0=P_{I_0}x_0,\qquad y_0=x_0-\w x_0,\qquad \w y_0=P_{I_0^c}y_0,\quad x_1=y_0-\w y_0.
\eaaa
\item
For $k\ge 1$, set
 \baaa
\w x_k=P_{I_k}x_k,\qquad y_k=x_k-\w x_k,\qquad \w y_{k}=P_{I_k^c}y_k,\quad x_{k+1}=y_k-\w y_k.
\eaaa
\end{itemize}
The following lemma will be useful.
\begin{lemma}\label{lemmaN}  For any $I\in\J$ and $x\in\ell_2^-$, the following holds:
\begin{enumerate}
\item $\|x\|\ge \|x-P_{I}x\|$, and  \item The equality in (i) holds if and only if $P_Ix=0$.
\end{enumerate}
\end{lemma}
\subsubsection*{Stopping upon arriving at a predictable process}
If there exists $k\ge 0$ such that $y_k=0$ then
\baa
x=\w x_0+y_0=\w x_0+\w y_0+x_1=\w x_0+\w y_0+\w x_1 +y_1=...=\w x_0+\w y_0+\w x_1 +\w y_1+...  +\w x_k.
\label{xx}\eaa
This means that $x$ is a finite sum of left band-limited processes. These processes were calculated by the observer,
and, in this sense, each of them can be deemed to be observed, with known (pres-selected) $I_k$;
in particular, $x$ can be predicted without error. Similarly, if there exists $k\ge 0 $ that  $x_{k+1}=0$, then
\baa
x=\w x_0+y_0=\w x_0+\w y_0+x_1=\w x_0+\w y_0+\w x_1 +y_1=...=\w x_0+\w y_0+\w x_1 +\w y_1+...  +\w y_k.
\label{xy}\eaa
 This means that $x$ again
 is a finite sum of observed  left band-limited processes. Again,  $x$ can be predicted without error. \par
The norms $\|\eta_k\|_{\ell_2^-}$ and $\|\oo \eta_k\|_{\ell_2^-}$ can be used for quantification of the
randomness of one-sided semi-infinite sequences.
 \subsubsection*{The case of never stopping procedure}
  It may happen that, for any $N>0$, there exists $k\ge N$ such that either
$\|y_k\|_{\ell_2^-}+\|x_k\|_{\ell_2^-}>0$. In this, the randomness can be quantified
as
\baaa
\max\left(\limsup_{k\to +\infty}\|x_k\|_{\ell_2^-},\limsup_{k\to +\infty}\|y_k\|_{\ell_2^-} \right).
\eaaa
 \subsubsection*{Arrival at a non-reducible noise}
 A process $x\in\ell_2^-$ is either left band-limited or not band-limited. Therefore, some processes cannot
 be represented as a finite sum of left bandlimited processes such as (\ref{xx}) or (\ref{xy}) with a finite $k$.
 In this case, the procedure will not be stopped according to the rule described above. It could be  beneficial
 to stop procedure using the following  rule.
\par
 Let \baaa
\d_k\defi \|x_k\|_{\ell_2^-}-\|x_k-\w x_k\|_{\ell_2^-},\qquad \oo\d_k\defi \|y_k\|_{\ell_2^-}-\|y_{k}-\w y_k\|_{\ell_2^-},
\eaaa
i.e.,
$\d_k=\|x_k\|_{\ell_2^-}-\|y_k\|_{\ell_2^-}$,  $\oo\d_k=\|y_k\|_{\ell_2^-}-\|x_{k+1}\|_{\ell_2^-}$,
 \baaa
\|x_{k}\|_{\ell_2^-}=\|y_k\|_{\ell_2^-}+\d_k=\|x_{k+1}\|_{\ell_2^-}+\d_k+\oo\d_k, \quad k=0,1,...\quad
\eaaa
\baaa
\|y_{k}\|_{\ell_2^-}=\|x_{k+1}\|_{\ell_2^-}+\oo\d_k=\|y_{k+1}\|_{\ell_2^-}+\d_k+\oo\d_k, \quad k=0,1,...\quad
\eaaa
 By Lemma \ref{lemmaN},
 it follows that $\d_k\ge 0$ and  $\oo\d_k\ge 0$ for all $k$, i.e., \baaa
\|x_{k}\|_{\ell_2^-}\ge \|y_k\|_{\ell_2^-}\ge \|x_{k+1}\|_{\ell_2^-}, \quad k=0,1,...\quad
\eaaa
By Theorem \ref{Th1n}, it may happen that $\d_k=0$, i.e., $\|x_{k}\|_{\ell_2^-}=\|y_k\|_{\ell_2^-}$.
To save the resources, the procedure should be stopped when this occurs, since further steps will not improve the result.
On this step, $x$ is presented as
\baaa
x=\w x_0+\w y_0+\w x_1 +\w y_1+... +\w x_k+ y_k=x_{p}^{(k)} + \eta_k.
\eaaa
where $x_{p}^{(k)}=\w x_0+\w y_0+\w x_1 +\w y_1+...+\w x_k$ is a predictable  process since it is a finite sum of observed
left band-limited processes, and $\eta_k=y_k$ is a noise. Given the selected set $\{I_k\}$, further reduction
of the norm of this noise is impossible. Hence  we can call $y_k$
a non-reducible noise.

Similarly, it may happen that  $\oo\d_k=0$ and $\d_k>0$, i.e. $\|y_{k}\|_{\ell_2^-}=\|x_{k+1}\|_{\ell_2^-}$.
Again, the procedure should be stopped when this occurs, since further steps will not improve the result.  This means that the procedure have
to stop on the step where $x$ is presented as
\baaa
x=\w x_0+\w y_0+\w x_1 +\w y_1+... +\w y_k+ x_{k+1}=y_{\BL}^{(k)} +\oo\eta_k.
\eaaa
Here $y_{p}^{(k)}=\w x_0+\w y_0+\w x_1 +\w y_1+... +\w y_k$ is a predictable process again, and $\oo\eta_k=x_{k+1}$
is a   non-reducible noise again.

\section{Proofs}\label{SceP}
 For the case where $I_0=(-\O,\O)$, i.e.
 $\o_I=0$, the proofs of Theorem \ref{lemmaPred}, Lemma \ref{propU} and  Theorems \ref{Th1}-\ref{ThP},  can be found in \cite{D12a}. Let us extend these proofs on case where $\o_I\neq 0$.

Let us observe that $x\in \XN$ and $X=\Z x\in \BN$ if and only if $x_0\defi p_{-\o_I}x\in \XNil(I_0)$ and
$X_0\defi \Z x_0\in \BNil(I_0)$. In this case, $x=p_{\o_I}x_0$, and
 \baaa X\ew=\sum_{t=-\infty}^{\infty}x(t)e^{-i\o t}=\sum_{t=-\infty}^{\infty}x_0(t)e^{i\o_I t}e^{-i\o t}= X_0\left(e^{i(\o_I -\o )t} \right), \quad
\o\in[0,2\pi). \eaaa
Then the proof of Theorem \ref{lemmaPred}(i)-(ii) and Lemma \ref{propU} follows.

\par
 {\em Proof of Theorem \ref{lemmaPred}} (iii) follows from the robustness of the predictor used in \cite{D12a} with respect to truncation
of inputs from $\ell_2$. $\Box$

Further, we have that
 \baaa  \|\w
x-x\|_{\ell_2^-}=\|p_{-\o_I}\w
x-p_{-\o_I} x\|_{\ell_2^-}  \quad\breakk\hbox{for any}\quad \w x, x\in \ell_2^-.\label{minn} \eaaa
Hence the problem
\baa &&\hbox{Minimize}\quad \|p_{-\o_I}\w
x-p_{-\o_I} x\|_{\ell_2^-} \quad\breakk\hbox{over}\quad \w x\in \XN \label{minnn} \eaa
has the  same sets of solution as problem (\ref{min}). Therefore, there is a bijection
between the sets  of optimal solutions for
problem (\ref{min}) and for the problem
\baa &&\hbox{Minimize}\quad \|\w
y-y\|_{\ell_2^-} \quad\breakk\hbox{over}\quad \w y\in \XNil(I_0),\label{minnnn} \eaa
where $y= p_{-\o_I} x $. This bijection has the form  $\w y= p_{-\o_I}\w x $.
Therefore, the proof for $\o_I\neq 0$ follows from the proof for $\o_I=0$ from \cite{D12a}.
Then the proof of Theorem \ref{Th1} and Theorem \ref{ThP} follows. $\Box$

{\em Proof of
Theorem \ref{ThO}}. \index{Let $m$ the entire part of the number $2\pi/\w\nu +1$, and let $\o_k=-\pi+2k\pi/m $, $k=1,...,m$.}
It is easy to see that there exists a finite set $\{I_k\}_{k=1}^M\subset \J$, $M<+\infty$, such that  $\mes(I_k)\le \nu/3$,
$\cup_{k=1}^M  I_k=(0,2\pi]$, and that the intersections of two different $ I_k$ cannot contain two or more elements.
Let $\w I_k=(-\pi,\pi]\setminus I_k$.

Let $P_I$ be operators such as defined in Section \ref{Sec1Sided}, with rather technical  adjustment: we assume that the
set of times
 $\{t\le 0\}$ in Theorem \ref{Th1} is replaced by  the $\{t\le \tau\}$,  and that $\ell_2^-$ replaced by $\ell_2(-\infty,\tau)$.
As is shown in Theorem \ref{ThP}, the values $d_k\defi\|\P_{\w I_k}x-x\|_{\ell_2(-\infty,\tau)}$ for $k=1,...,M$ can be found
based on observations of $\{x(t)\}_{t\le\tau}$. By the assumptions on $x$, there
exists $m$ such that
$d_m=0$. The  set $\w I=\w I_m$ is such as described in the Theorem; the point $\w\o_c$ can be defined as
select  $\w\o_{c}=\w\o_{\w I}-\pi$ if $\w\o_{\w I}\in (0,\pi]$, and $\w\o_{c}=\w\o_{\w I}+\pi$ if $\w\o_{\w I}\in (-\pi,0]$.
Then the proof of Theorem \ref{ThO} follows. $\Box$
 $\Box$

{\em Proof of Theorem \ref{Threc}}.
Let $Y\ew = \sum_{k\in\ZZ\setminus\{m\} } e^{-i\o k}x(k)$, $\o\in(-\pi,\pi]$; this function to be observable. By the definitions, it follows that
 \baaa
 X\ew -Y\ew -e^{-im}x(m)\equiv 0,\quad \o\in(-\pi,\pi].
 \eaaa
Hence
\baaa
x(m)=-e^{ im}Y\left(e^{\om}\right)+ e^{ im}X\left(e^{\om}\right)=-e^{ im}Y\left(e^{\om}\right)+\xi,
 \eaaa
where $\xi=e^{ im}X\left(e^{\om}\right)$. Hence
\baaa
|x(m)+e^{ im}Y\left(e^{\om}\right)|=  |\xi|=\s.
\eaaa
Let us accept the value $\w x(m)=-e^{ im}Y\left(e^{\om}\right)$ as the estimate of the
missing value $x(m)$. For this estimator, the size of the recovery error is $\s$ for any $x\in\X_\s$.
If $\s=0$ then the estimator is error-free.  In a general case where $\s\ge 0$, we have that (\ref{opt}) holds.

Let us show that  this estimator
is optimal in the following sense:
\baaa
\s=\sup_{x\in\X_\s}|\w x(m)-x(m)|\le \sup_{x\in\X_\s}|\ww x(m)-x(m)|
\label{optrec2}\eaaa
for any other  estimator $\ww x(m)=F\left(x|_{t\in\ZZ\setminus\{m\}}\right)$, where $F:\ell_2(-\infty,m-1)\times \ell_2(m+1,+\infty)\to\R$ is some mapping.

\index{Assume that $m=0$, $X_\pm\ew =\pm \s$, $x_\pm=\Z^{-1}X_\pm$, i.e. $x_\pm(t)=\pm \s\Ind_{\{t\neq 0\}}$.
Clearly, $x_\pm\in\X_\s$ and $\ww x_-=\ww x_+$ for $\ww x_\pm= F\left(x|_{t\in\Z,\ t\neq 0}\right)$, for any mapping $F$ such as described above.
Hence
\baaa
 \max(|\ww x_-(0)-x_-(0)|,|\ww x_+(0)-x_+(0)|)\ge \s.
\eaaa}

Let  $m\in \ZZ$ be fixed, and let  $X_\pm\ew =\pm \s e^{-im \o}$, $x_\pm=\Z^{-1}X_\pm$, i.e. $x_\pm(t)=\pm \s\Ind_{\{t=m\}}$.
Clearly, $x_\pm\in\X_\s$. Moreover, we have that $\ww x_-=\ww x_+$ for $\ww x_\pm= F\left(x|_{t\in\ZZ\setminus\{m\}}\right)$, for any mapping $F$ such as described above.
Hence
\baaa
 \max(|\ww x_-(m)-x_-(m)|,|\ww x_+(m)-x_+(m)|)\ge \s.
\eaaa
 Then (\ref{optrec}) follows. This completes the proof of Theorem \ref{Threc}. $\Box$

\par
{\em Proof of Theorem \ref{Th1n}}. It suffices to observe that $\ell_2^-\setminus \Q(\ell_2)\neq \emptyset$, for the operator $\Q:\ell_2\to \ell_2^-$, since
$\Q(\ell_2)=\ell_2^{-,\LBL}$.
 Hence the kernel of the adjoint operator   $\Q^*:\ell_2^-\to\ell_2$ contains non-zero elements.
 \index{Since $R$ is invertible, we have that
$R^{-1}\Q^*x^\bot\neq 0_{\ell_2}$. Clearly, $\Q y\neq 0_{\ell_2^-}$ if $y\neq 0_{\ell_2}$. Hence  $\Q R^{-1}\Q^*x^\bot\neq 0_{\ell_2^-}}
$\Box$

\par
{\em Proof of Lemma \ref{lemmaN}}. Statement (i) follows from the choice of $P_{I}x$ as a solution of
optimization problem (\ref{min}). To prove statement (ii), it suffices to show that if $\|x\|=\|x-P_{I}x\|$ then $P_Ix=0$.
If $\|x\|=\|x-P_{I}x\|$ then  $\|x-0_{\ell_2^-}\|=\|x-P_{I}x\|$. Hence both sequences $0_{\ell_2^-}$ and $\|P_{I}x\|$
are solutions of problem (\ref{min}). We proved that the solution is unique, hence $\|P_{I}x\|=0_{\ell_2^-}$.
This completes the proof. $\Box$
\section{Possible applications and future development}
The  approach suggested in this paper  allows many modifications. We
outline below some possible straightforward modifications as well as
  more challenging problems and possible applications  that we leave for the future research.
\begin{enumerate}
\item It would be interesting to investigate sensitivity of the prediction results with respect to the choice
of
$\{I_k\}$. It would be interesting to find  an optimal choice of the set $\{I_k\}$ such as
    \baaa
    \hbox{Maximize}\quad  \d_k+\oo \d_k\quad \hbox{over}\quad  I\in \J
    \eaaa for $k=1,2,..,$, with some constraints on the choice of $I_k$, for example, such that $\mes(I_k)$ is given.
\item It could be interesting  to try another basis in $L_2(I_0)$ for expansion in (\ref{wX}).
\item Optimization problem in (\ref{min}) is based on optimal approximation in
$L_2(I)$ for Z-transforms. This approximation in  can be replaced by approximation in a  weighted
$L_2$-space on $I$. This leads to modification of the optimization
problem; the weight will represent the relative importance of the
approximation on different frequencies.
\item It is unclear if an analog of property (\ref{N}) can be obtained with $d=2$ instead of $d=1,+\infty$.
\end{enumerate}

\end{document}